# Development of an alternative PSO-based algorithm for simulation of Endurance Time Excitation Functions

Mohammadreza Mashayekhi[a*], Mojtaba Harati[b*], Homayoon E. Estekanchi[c]

a) Research associate, Department of Civil Engineering, Sharif University of Technology. E-mail: mmashayekhi67@gmail.com

b) Corresponding author, Lecturer, Department of Civil Engineering, University of Science and Culture. E-mail: moj.harati@gmail.com

c) Professor, Department of Civil Engineering, Sharif University of Technology. E-mail: stkanchi@sharif.edu

**Abstract**

This paper presents a particle swarm optimizer for production of endurance time excitation functions. These excitations are intensifying acceleration time histories that are used as input motions in endurance time method. The accuracy of the endurance time methods heavily depends on the accuracy of endurance time excitations. Unconstrained nonlinear optimization is employed to simulate these excitations. Particle swarm optimization method as an evolutionary algorithm is examined in this paper to achieve a more accurate endurance time excitation function, where optimal parameters of the particle swarm optimization are first determined using a parametric study on the involved variables. The proposed method is verified and compared with the trust-region-reflective method as a classical optimization method and imperialist competitive algorithm as a recently developed evolutionary method. Results show that the proposed method leads to more accurate endurance time excitations.

**Key words**: endurance time method; particle swarm optimization; dynamic analysis; discrete wavelet transform; classical optimization methods, imperialist competitive algorithm.

## Nomenclature

| | |
|---|---|
| $a_g(\tau)$ | acceleration time history of an Endurance Time excitation |
| ET | endurance time |
| ETEF | endurance time excitation function |
| $g(t)$ | a function pertinent to the intensifying profile of an ETEF |

---

[*]These authors have contributed equally to the work

| | |
|---|---|
| ICA | imperialist competitive algorithm |
| $m$ | sample number of periods |
| $n$ | sample amount of time |
| $n_{pop}$ | size of a selected population |
| $N_{var}$ | dimension of an optimization problem |
| PSO | particle swarm optimization |
| $S_1$ | a parameter related to the long-period spectral response acceleration |
| $S_a(t,T)$ | acceleration spectra produced by ETEFs at time $t$ and period $T$ |
| $S_a^{target}(T)$ | target acceleration spectra |
| $S_S$ | a parameter for a short-period spectral response acceleration |
| $T_{max}$ | maximum period of vibration in the simulation procedure |
| $t_{max}$ | motion duration of an ETEF |
| $T_{min}$ | minimum period of vibration in the simulation procedure |
| $x_i$ | an optimization variable |
| $\ddot{x}(\tau)$ | acceleration response of an SDOF |
| $x_{i,max}$ | an upper bound for optimization variables |
| $x_{i,min}$ | a lower bound for the optimization variables |
| $\xi$ | a recursive linear function for inertia weight in PSO |
| $k$ | a pseudo-time increment in PSO algorithm |
| $V_i^k$ | current position of the i-th particle |
| $X_i^k$ | velocity of the i-th particle |
| $P_i^k$ | the best previous position of the i-th particle at time $k$ |
| $P_g^k$ | the best global position of the swarm at time $k$ |
| $r_1, r_2$ | two uniform random sequences generated uniformly between 0 and 1 |
| $\omega$ | inertia weight used to discount previous velocity of the particle |
| $c_1, c_2$ | constant factors in PSO algorithm |
| $K$ | a constant multiplier in constriction coefficient approach |
| CCA | constriction coefficient approach |

# 1 Introduction

In recent years, frameworks such as ASCE07 [1] and rehabilitation provisions (e.g. ASCE/SEI 41-17 [2] and FEMA-356 [3] demand an accurate estimation of seismic responses very often [4,5]. As a result, nonlinear dynamic procedures should be regarded as a potent candidate framework to compute structural demands under earthquake seismic inputs. This is based on the fact that such dynamic procedures are capable of incorporating nonlinearities existed in both structural and material levels. Endurance Time (ET) method is a type of such dynamic time history analyses in which structures are subjected to intensifying predefined acceleration motions [6], offering seismic demand prediction of structures in terms of the correlation between engineering demand parameters and intensity measures. The primary advantage of the endurance time (ET) method over the conventional time history analysis—which makes use of recorded ground motions—is that a considerable reduction in the required computational time is secured once ET is employed. When the ET method is compared to other existing incremental-based dynamic procedures [7,8], this feature more facilitates the required efforts to execute nonlinear assessment especially for the practical use in engineering offices. This is mainly due to the fact that there is no need to scale up and down of the selected recordset required to perform such incremental-based dynamic analyses. Besides, the difficulties associated with the record selection procedure in such incremental-based dynamic analyses [9,10] can be also circumvented because three couples of ET records would be enough seismic assessment within the framework of Endurance Time Analysis (ETA). The ET method has been extensively used in different areas of earthquake engineering, such as those related to the seismic assessment RC and steel structures [11–13], seismic performance of bridges [14], and duration-consistent seismic evaluations [15,16].

In the ET method, structures are subjected to intensifying acceleration time histories which are also called as the endurance time excitation functions (ETEF). The more accurate the ETEFs are, the more reliable the ET method's outputs will be [6]. ETEFs are artificial acceleration time histories that are generated mathematically. ETEFs are intensified with time while they preserve their compatibility with recorded ground motions. Having more accurate ETEFs is demanded in the successful implementation of the ET method, emphasizing the importance and efficiency of the methods used to simulate ETEFs [6]. Several researches have been aimed to increase the efficiency of the simulation methods for ETEFs. In the simulation procedure, the objective is to minimize the discrepancy between ETEFs and real ground motions [17,18]. Thus, equations are expressed in order to account for the discrepancies of ETEFs and real ground motions. Analytical solution for the mentioned equations does not exist because the number of equations is considerably more than the number of variables [15]. As a result, an optimization procedure is adopted to solve the equations. In the optimization context, these equations are expressed in term of objective functions.

As mentioned before, simulating ETEFs is an optimization problem that intends to minimize a predefined objective function. The dynamic nature and high number of optimization variables differentiate this problem from other conventional optimization problems [17]. These two issues lead to the presence of many local optima in the problem. Consequently, solving this problem requires a method that appropriately deals with the mentioned difficulties.

In the field of simulating ETEFs, the number of studies that focused on the development of appropriate objective functions and optimization spaces is appreciably more than those studies attempted to improve the employed solution methods. The study by Mashayekhi et al. [15] that modified the objective function for simulating ETEFs to include cumulative absolute velocity (CAV), and the study by Mashayekhi et al. [16] that developed hysteretic energy (HE) compatible objective functions are examples of the studies aimed to improve objective functions of ETEFs. Given the fact that motion duration can have a significant influence on the structural responses [19–22], these above-mentioned studies tried to improve objective functions of ETEFs by including two prominent duration-related parameters—the CAV and EH—for such simulations.

In contrast to local optimizers such as classical methods that find optimal solutions in the vicinity of starting points, global optimizers such as evolutionary algorithms search the whole optimization space regardless of the initially selected starting points. Although using evolutionary algorithms seem to be justifiable than the classical methods due to the fact that they do not get trapped in local optima, existing ETEFs are yet generated by classical optimization methods [16]. Besides, simulating ETEFs by classical optimization methods is not straightforward and requires several trials and errors to find the best potential solution because optimization results are very sensitive to initial points [17]. So, this issue complicates the simulation process of ETEFs. On the other hand, a high number of optimization variables and the complexity of ETEFs objective functions are obstacles to employ evolutionary algorithms in simulating ETEFs. In this case, these issues have to be perfectly addressed when using evolutionary algorithms in simulating ETEFs.

Recently, many evolutionary algorithms which are based on computer simulation of the natural process have become more attractive. These algorithms have been widely used in different engineering practice. In this case, Mashayekhi et al. [17] employed the imperialist competitive algorithm (ICA) [23] in simulating ETEFs. Although ICA proposed by Atashpaz-Garagari [23] has been used to simulate endurance time excitations, other developed evolutionary algorithms have not been examined to check out whether or not they generate more accurate ETEFs. The Particle Swarm Optimization (PSO) is another technique motivated by the social behaviors of bird flocking and fish schooling [24]. Compared to other evolutionary algorithms such as ICA, the PSO has fewer parameters and easy to implement for a simulation-based investigation. It is worth to add that evolutionary algorithms have several parameters that should be specified in advance, where finding optimal parameters of these evolutionary algorithms for a simulation-based investigation is a separate optimization problem. Thus, this procedure could be more complicated if the employed evolutionary algorithm has a large number of parameters.

This paper presents a PSO-based algorithm to simulate ETEFs. Various parameters of PSO are first examined to find the optimal parameters. Then the proposed method for simulating ETEFs is applied to a case study. The accuracy of newly generated ETEFs is examined by comparing their dynamic characteristics with the targets. Additionally, the obtained results of the proposed method are compared with two algorithms that have already been employed for simulating ETEFs—namely trust-region reflective [25] and ICA-based method. The trust-region reflective is a gradient-based mathematical algorithm by which existing ETEFs have been generated. The ICA is an evolutionary algorithm that has been recently employed for simulating ETEFs. The accuracy and the required computational time of the proposed method are compared with the mentioned algorithm, the ICA-based method.

## 2 Simulation of Endurance Time Excitation Functions

The production of ETEFs is finding appropriate values for decision variables so that the minimum mismatch with targets is achieved. There are different ways to define decision variables in this problem. This study employs reduced discrete wavelet transform (DWT) space to represent ETEFs. The wavelet analysis is been used in many fields. Recent applications of the wavelet transform in the field of civil engineering are readily found in several types of research, including the one related to the dynamic analysis of structures [26], damage detection [27,28] and system identification [29]. Readers are referred to Mashayekhi et al. [30] for more details on finding how DWT can be employed for simulation of new ETEF functions. In this study, the first eight levels of the wavelet are considered, therefore, decision variables calreduce to 512.

ETEFs objective functions computes mismatch of ETEFs dynamic characteristics and targets. Intensities of ETEFs have to increase with time and be compatible with a suite of ground motions. Acceleration spectra are used as intensity measures in producing endurance time excitations. Acceleration spectra of ETEFs are expressed as shown by Equation (2) in order to satisfy the mentioned requirements. In this equation, Subscript "$T$" denotes target. $S_{aT}(t,T)$ represents target acceleration response spectra of ETEFs.

$$S_{aT}(t,T) = g(t) * S_a^{\text{target}}(T) \tag{2}$$

where $S_a^{\text{target}}(T)$ is taken as the target acceleration response spectrum considered to be a work at the target time. And $g(t)$ is a selected intensifying function, which is a function of time itself and can control the profile shape of the increasing acceleration time history of ETEFs. The term $S_a(t,T)$ also stands for the acceleration response spectra of the ETEFs, which can be generated at the time $t$ and period $T$.

Acceleration response spectra of the ETEFs, at time $t$ and period $T$, can be readily computed using Equation (3):

$$S_a(t,T) = \max\left(\left|\ddot{x}(\tau) + a_g(\tau)\right|\right) \quad 0 \leq \tau \leq t \tag{3}$$

In the above equation, $\ddot{x}(\tau)$ is the relative acceleration response of an SDOF system under the seismic input provided by an ETEF when the SDOF has a natural period of vibration $T$ and a damping ratio equal to 5%.

The objective function of simulating linear ETEFs through which only acceleration spectra are included for simulation iterations is brought in Equation (4). It should be mentioned that integration is performed over all periods and times.

$$F_{\text{ETEF}}(a_g) = \int_{T_{\min}}^{T_{\max}} \int_0^{t_{\max}} \left\{\left[S_a(T,t) - S_{aT}(T,t)\right]^2\right\} \mathrm{d}t \mathrm{d}T \tag{4}$$

where $t_{\max}$ is the excitations duration length while $T_{\min}$ and $T_{\max}$ are minimum and maximum periods considered in the simulation procedure.

In this study, design spectra proposed by ASCE07 standard [1] is used as target spectra. The study performed by Mirzaee and Estekanchi [6] is used in order to obtain the essential parameters in this case: $S_1$ and $S_s$ for Tehran. These parameters are respectively long and short spectral acceleration parameters.

Discretization should be carried out for numerically solving the objective function, and consequently, the type of discretization has influence on the final results. If time duration *t* is sampled at *n* points $t_j(j=1:n)$ and periods is sampled at *m* points $T_i(i=1:m)$, the objective function of Equation (4) converts to the following Equation:

$$F_{\text{ETEF}}(a_g) = \sum_{i=1}^{m}\sum_{j=1}^{n}\left\{\left[S_a(T_i,t_j) - S_{aC}(T_i,t_j)\right]^2\right\} \tag{5}$$

In this study, the period is sampled at 120 points that are logarithmically distributed between 0.02 and 5. With the logarithmic distribution, more data in the low period region are produced. The fluctuation of acceleration spectra in the low period zone is indeed high. Time variable *t* is also sampled at 2048 points with equal intervals of 0.01 second.

## 3   Particle swarm optimizer (PSO)

The particle swarm optimization (PSO) has been inspired by the social attitude of birds [31], and it is a population-based optimization. Its population is called a *swarm* and each individual is called a *particle*. Each particle is a potential solution to the optimization problem. Each particle flies through optimization space to search for optima. Particles possess three general attributes: 1) A current position that represents a potential solution, 2) A current velocity that controls its fly speed and direction, 3) An objective function value that determines its merit. Particles exchange information about their position, velocity, and the objective function value and thus increasing the probability of migration to regions with low objective function values. Joint cooperation between particles is the distinguishing feature of the PSO framework.

The PSO starts with a swarm consisting of a number of particles, which are randomly generated in the search space of the objective function. Particles fly through the search space by the help of their velocities. Velocities that determine particles flying direction are obtained for each particle based on its previous best position and the characteristics of a particle with the best position in the whole swarm. The best position of the swarm is the corresponding position of a particle that has the minimum objective function value among all particles in the swarm. This strategy for calculating velocities increases the probability of migration of the particles to regions with the lower objective function. Particle positions are changed after each flight and the corresponding objective function value of the particles are evaluated for updated positions. A schematic flight of a particle in the PSO is shown in Figure 1.

For particle *i,* in each iteration, the previous position of the particle is changed according to the following equations:

$$\begin{aligned}V_i^{k+1} &= \omega V_i^k + c_1 r_1\left(P_i^k - X_i^k\right) + c_2 r_2\left(P_g^k - X_i^k\right) \\ X_{i+1}^k &= X_i^k + V_i^{k+1}\end{aligned} \tag{6}$$

where Subscript *k* indicates a pseudo-time increment; $V_i^k$ and $X_i^k$ represent the current position and the velocity of the i-th particle; $P_i^k$ is the best previous position of the i-th particle at time *k* (called *pbest*), and $P_g^k$ is the best global position of the swarm at time *k* (called *gbest*); $r_1$ and $r_2$ are two uniform random sequences generated uniformly between 0 and 1; $c_1$ and $c_2$ are constants in PSO algorithm; $\omega$ is the inertia weight used to discount previous velocity of the particle preserved. A larger inertia weight makes the global exploration easier while a smaller inertia weight brings about local exploration [20]. At the end of the run, the algorithm may get trapped in the local minimum. By using the linearly decreasing inertia weight, the PSO lacks global search ability at the end of run to solve the mentioned matter. In this study, two alternatives for inertia weight tuning are considered. First, a constant value is assigned to inertia weight for the entire iterations within the algorithm. Second, inertia weight is decreased with a recursive linear function as expressed below:

$$\omega^k = \xi \times \omega^{k-1} \qquad (7)$$

where $\xi$ is a damping factor. The initial value of inertia weight is set to 1. Different values for damping factors are examined in this study.

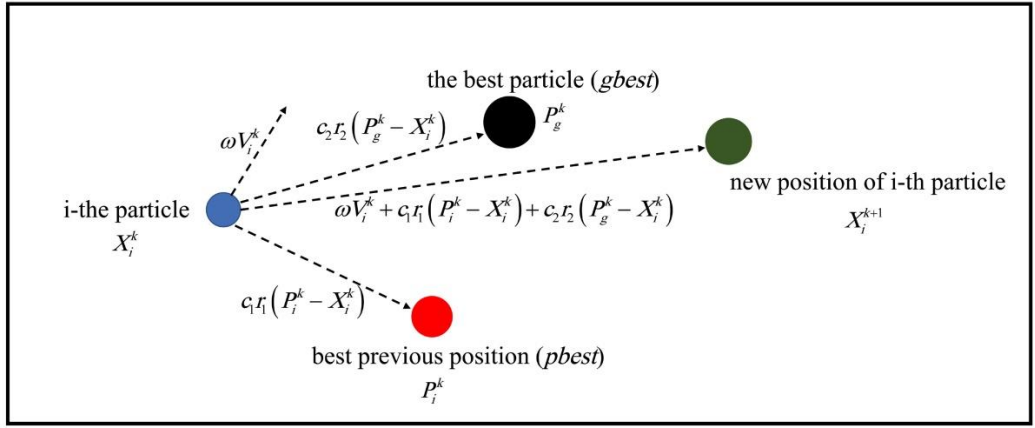

Figure 1. The schematic illustration of a flight in the PSO

## 4  Constraint handling method

ETEFs simulating problem is, in general, an unconstrained nonlinear optimization problem. In this paper, a lower and upper bound on variables are prescribed to increase the performance of the algorithm. Limiting the search space removes unwanted particles from the swarm, and therefore, improves the efficiency of the algorithm. Unwanted particles refer to those particles that have a very low chance and potentiality to become a solution for the problem. Defining upper and lower bounds for variable implies that inequality constraints according to Equation (8) are considered and no equality constraint is included in the problem.

$$x_{i,\min} \leq x_i \leq x_{i,\max} \qquad 1 \leq i \leq N_{\text{var}} \qquad (8)$$

where $x_{i,\min}$ and $x_{i,\max}$ are lower and upper bounds of the decision variable $x_i$. The above equation has two inequality constraints. It can be deduced that the total constraint number of an ETEF simulating problem is as twice the number of decision variables.

However, some experimental results indicate that this technique (the penalty function) reduces the efficiency of the PSO because it returns the infeasible particles to their previous best positions (*pbest*), through which a search mechanism from reaching out to a global minimum would be prevented and inactivated [32]. One more drawback of using penalty functions is that they require additional tuning parameters. There is another constraint handling method called "fly-back" that can be employed in such a PSO-based optimization problem. In this method and as can be seen in Figure 2, when a particle flies to outside of the feasible region, it flies back to previous position. For a good review on the existing techniques concerning the constraint-handling mechanism, readers are referred to a rather complete study by Coello and Carlos [33].

In this paper, a new approach inspired by the method of fly-back is used to handle the constraints. In the proposed method, if a particle does not comply with variable bounds, that particle flies back to the previous position and then flies to a new position by using a new random number. It is also checked whether the new position is an infeasible region or not. The key advantage of the proposed method is its simplicity and that it does not need any further tuning parameter. The main difference between the proposed technique and the fly-back method is that fly-back method searches for a new position for the violated particle at the next iteration, but the proposed method carries out the search at the current iteration with a new random number.

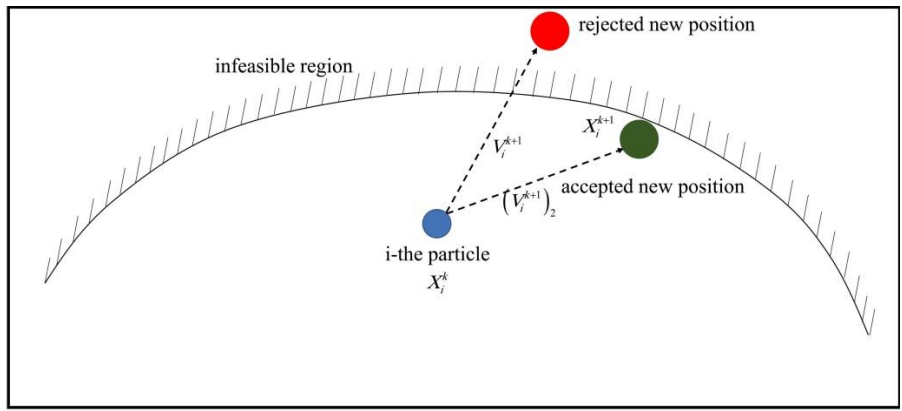

Figure 2. The illustration of the proposed constraint handling in the PSO

## 5 Algorithm

In this section, the implementation outline of the proposed PSO-based simulating ETEFs procedure is presented. The pseudo-code of the algorithm is as follows.

**Step 1**: Generating initial swarm

PSO starts with an initial population called "swarm". Each swarm contains several particles. Each particle is identified by a set of decision variables as following:

$$particle_i = \left[ x_i, x_2, ..., x_{N_{var}} \right] \tag{9}$$

wheer $N_{var}$ is the dimension of the optimization problem.

Primary locations of particles are found using a group of values that are assigned to each decision variable. The type of assigning these initial particles must ensure that these particles cover the entire search space. In order to satisfy the mentioned condition, two issues must be carefully considered. First, enough number of particles must be opted. Second, type of random number generation should be selected in a way that each possible value for the optimum solution is created.

As a result of ETEFs dynamic nature, decision variables of an ETEF simulation problem are highly correlated. In order to consider this correlation in generation of the initial swarm, the method developed by [17] is employed. It is of noteworthy to recall that the method described above is not restricted to solve the specific problem of ETEFs simulation, so it may be readily hired for other problems too.

**Step 2**: Evaluating objective function values of the particles

Generated particles in step 01 are expected to cover the entire optimization space, each of which lie in different positions. The objective function values of the particles are evaluated, where the value related to an objective function shows the merit of a particle.

**Step 3**: Sorting the swarm and calculating the best particle

The best particle that has the minimum objective function value is identified here. The best particle is employed for updating positions of the other particles. In other words, particles share their information by using the observed characteristics of the best particle.

**Step 4**: Moving particles toward new positions

The positions of the particles are changed according to Equation (6). All changes are accepted unless the new position lies outside the variable ranges [$x_{min}$, $x_{max}$]. If this condition happens for a particle, Equation (6) with a new random number is reiterated. This procedure is repeated until the new position is acceptable.

**Step 5**: Updating the best particle

The best particle of new population is found and compared with the best particle found in previous stages. If the former has a less objective function value than the latter, it is considered as the best particle; otherwise, the present best particle is not replaced with a new one.

**Step 6**: Terminating criteria control

There are numerous stopping criteria which can be used. In this study, when the number of iterations reaches a pre-defined value, the algorithm is stopped. The best particle ever found in the process is considered as the solution.

A summary of the PSO algorithm which is implemented in this study is depicted in Figure 3.

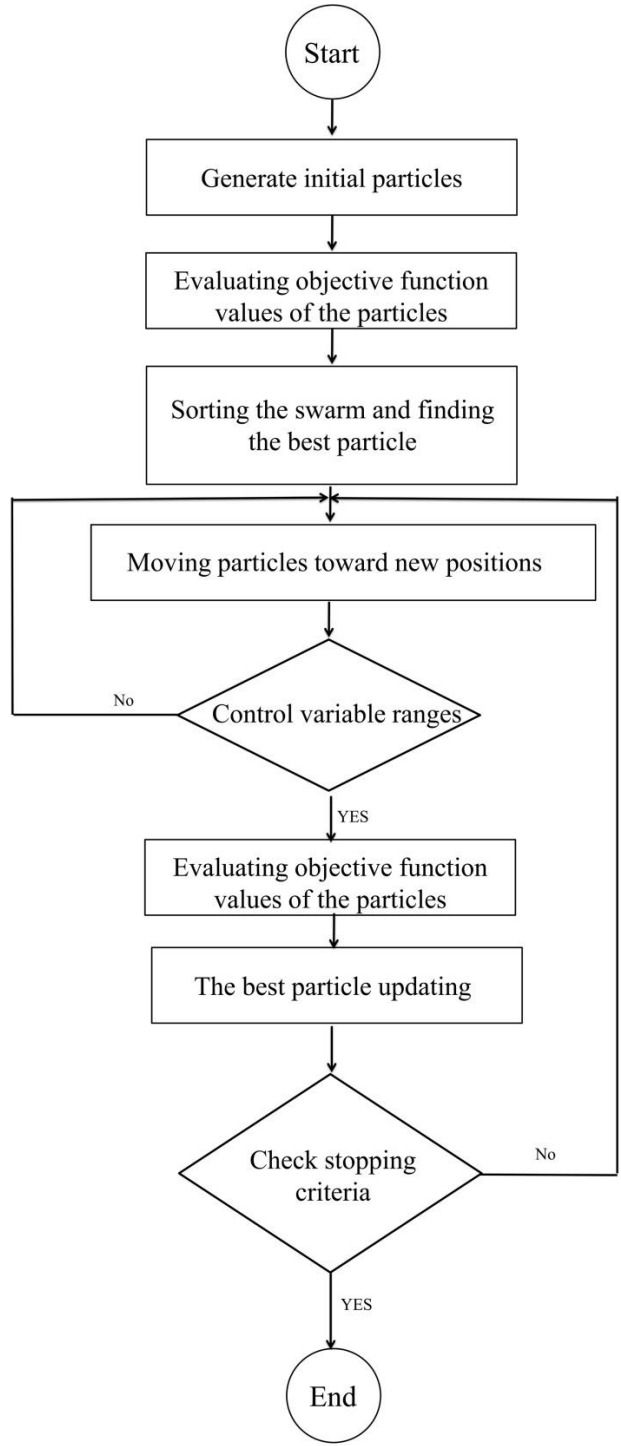

Figure 3. The schematic framework for the PSO algorithm used for ETEFs simulation

## 6 Optimal parameters of the proposed PSO framework

In this section, the configured algorithm to study the efficiency of a PSO-based method is applied to the specific problem of generating or simulating the new ETEFs. But before the simulation phase, we have to obtain the required optimal parameters of the PSO in this regard:

the w, c1 and c2. In this case, 28 optimization scenarios all of which are capable to simulate new ETEF would be considered here for some parametric studies. The specifications related to each of selected scenarios are presented in Table 1. For these scenarios, the constriction coefficient approach (CCA) is also included for a number of cases—ET_PS25 to ET_PS28 in Table 1. In this approach developed by Clerc and Kennedy [34], the constriction factor is used to guarantee convergence of the particle swarm optimization. The velocity of CCA is expressed as follow:

$$V_i^{k+1} = K\left[V_i^k + c_1 r_1 \left(P_i^k - X_i^k\right) + c_2 r_2 \left(P_g^k - X_i^k\right)\right] \qquad (10)$$

where $K$ is a constant multiplier computed as follow:

$$K = \frac{2}{\left|2 - \varphi - \sqrt{\varphi^2 - 4\varphi}\right|}, \text{ where } \varphi = c_1 + c_2, \varphi > 4 \qquad (11)$$

The convergence characteristic of the system is controlled by $\varphi$. In CCA, $\varphi$ must be greater than 4 to ensure convergence. When the CCA is used, $\varphi$ is set to 4.1 (i.e. $C_1 = C_2 = 2.05$) and constant multiplier K is thus 0.729. This leads to the previous velocity being multiplied by 0.729 and the terms $(Pbest_i^k - X_i^k)$ and $(Gbest^k - X_i^k)$ being multiplied by $2.05 \times 0.729 = 1.49445$. In contrast to other evolutionary computation methods, the CCA is based on the mathematical theory.

In order to obtain the above-mentioned optimal values of this devised PSO-based algorithm, the pertinent best value of the cost function in each scenario is sought to be calculated using the PSO-based method recommended in this paper. The results obtained from this parametric study are given in Table 1, where the best cost value for each scenario is located on the right-hand side of the table. It is worth to note that a mean and a standard deviation of 1966.13 and 1134.85 have been found, respectively, for the minimum objective values of the optimization scenarios executed here in this section. As can be witnessed for the data available in Table 1, there is considerable variability in all the objective values. In this case, a COV of 58% is seen for the data of cost values in Table 1, which is deemed to be high altogether. Hence it is a must to work with a set of best parameters, which are also optimum on a case-by-case basis, in this PSO-based algorithm for the simulation of new ETEFs.

A combination of the parameters, including ω=0.8, $\omega_p = 1$ (damping parameter) and $C_1 = C_2 = 1$, leads to more accurate simulated ETEFs in all cases. In this case, the above-mentioned combination of parameters can deliver an objective function of 737.62 and 787.83 for the number of population ($n_{pop}$) of 400 and 800, respectively. So, a population size of 400 is selected hereafter for the simulation of ETEFs. Optimal parameters of this proposed PSO-based method reported in Table 2, where the number of values to control this parametric study for each selected variable is also incorporated in this table.

Table 1. Specifications used to perform a parametric study on the variables involved in the proposed simulation procedure

| Scenario ID | $n_{pop}$ | Restriction factors | $\omega$ | $\omega_p$ | $c_1$ | $c_2$ | Minimum objective function values |
|---|---|---|---|---|---|---|---|
| ET_PS01 | 400 | 1 | 1 | 1 | 2 | 2 | 4079.26 |
| ET_PS02 | 400 | 1 | 1 | 0.99 | 2 | 2 | 2177.84 |
| ET_PS03 | 400 | 1 | 0.8 | 1 | 2 | 2 | 4023.87 |
| ET_PS04 | 400 | 1 | 0.8 | 0.99 | 2 | 2 | 1747.42 |
| ET_PS05 | 400 | 1 | 1 | 1 | 1 | 1 | 3446.56 |
| ET_PS06 | 400 | 1 | 1 | 0.99 | 1 | 1 | 1407.08 |
| ET_PS07 | 400 | 1 | 0.8 | 1 | 1 | 1 | 737.62 |
| ET_PS08 | 400 | 1 | 0.8 | 0.99 | 1 | 1 | 1280.56 |
| ET_PS09 | 400 | 1 | 1 | 1 | 0.5 | 0.5 | 3332.59 |
| ET_PS10 | 400 | 1 | 1 | 0.99 | 0.5 | 0.5 | 1752.80 |
| ET_PS11 | 400 | 1 | 0.8 | 1 | 0.5 | 0.5 | 857.42 |
| ET_PS12 | 400 | 1 | 0.8 | 0.99 | 0.5 | 0.5 | 1722.08 |
| ET_PS13 | 800 | 1 | 1 | 1 | 2 | 2 | 3704.65 |
| ET_PS14 | 800 | 1 | 1 | 0.99 | 2 | 2 | 1596.35 |
| ET_PS15 | 800 | 1 | 0.8 | 1 | 2 | 2 | 3734.29 |
| ET_PS16 | 800 | 1 | 0.8 | 0.99 | 2 | 2 | 1804.74 |
| ET_PS17 | 800 | 1 | 1 | 1 | 1 | 1 | 3662.53 |
| ET_PS18 | 800 | 1 | 1 | 0.99 | 1 | 1 | 1198.36 |
| ET_PS19 | 800 | 1 | 0.8 | 1 | 1 | 1 | 787.83 |
| ET_PS20 | 800 | 1 | 0.8 | 0.99 | 1 | 1 | 1213.02 |
| ET_PS21 | 800 | 1 | 1 | 1 | 0.5 | 0.5 | 2867.24 |
| ET_PS22 | 800 | 1 | 1 | 0.99 | 0.5 | 0.5 | 1566.26 |
| ET_PS23 | 800 | 1 | 0.8 | 1 | 0.5 | 0.5 | 872.95 |
| ET_PS24 | 800 | 1 | 0.8 | 0.99 | 0.5 | 0.5 | 1755.70 |
| ET_PS25 | 400 | CCA | Eq. (11) | 1 | 2.05 | 2.05 | 848.16 |
| ET_PS26 | 400 | CCA | Eq. (11) | 0.99 | 2.05 | 2.05 | 1113.96 |
| ET_PS27 | 800 | CCA | Eq. (11) | 1 | 2.05 | 2.05 | 784.68 |
| ET_PS28 | 800 | CCA | Eq. (11) | 0.99 | 2.05 | 2.05 | 976.06 |

Table 2. Optimal values found for the variables related to the PSO-based simulation framework

| Examined parameter | The value found to be optimum | Total number of considered values |
|---|---|---|
| $n_{pop}$ | 400 | 2 |
| Restriction factor | 1 | 1 and CCA |
| $\omega$ | 0.8 | 2 |
| $\omega_p$ | 1 | 2 |
| $c_1$ | 1 | 4 |
| $c_2$ | 1 | 4 |

## 7 Results

In this section, results associated with the time histories as well as the acceleration spectra of the simulated ETEFs are presented. The time history of one of those ETEFs produced using the PSO-based method is shown in Figure 4. As can be seen, it is an intensifying earthquake motion and has an increasing trend as it was devised to be in the simulation process. These generated ET records are called ETEF-PSO hereafter.

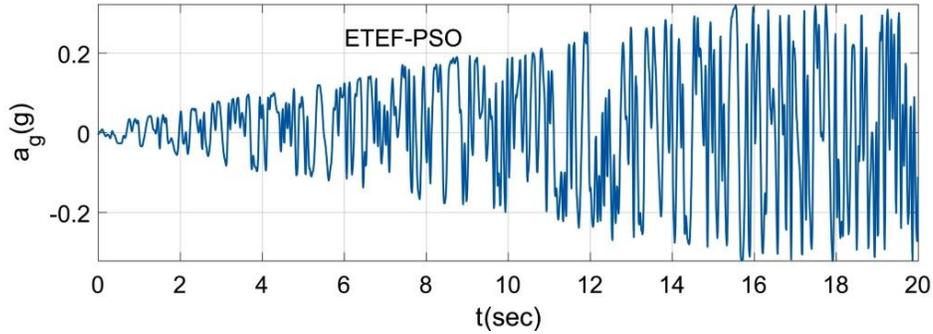

Figure 4. The ET records simulated from the proposed PSO framework, the ETEF-PSO

Furthermore, the response spectra obtained from these ETEFs are compared with the target spectra employed for the simulation of ET excitations. In Figure 5, these response spectra resulted from ETEFs are visually compared with the ones considered as the target spectra at time $t$=5sec, 10sec, 15sec, and 20sec. As can be found from this figure, it is apparently recognizable that the simulated spectra and the targets are in good agreements with each other. Similarly, comparisons are also made between the time variation of response spectra—resulted from ETEF-PSO excitations—and the targets for different periods of T=0.05sec, 0.8 sec, 2.0 sec, and 3.0 sec. We can observe an appropriate compatibility of targets with the time variation of simulated response spectra.

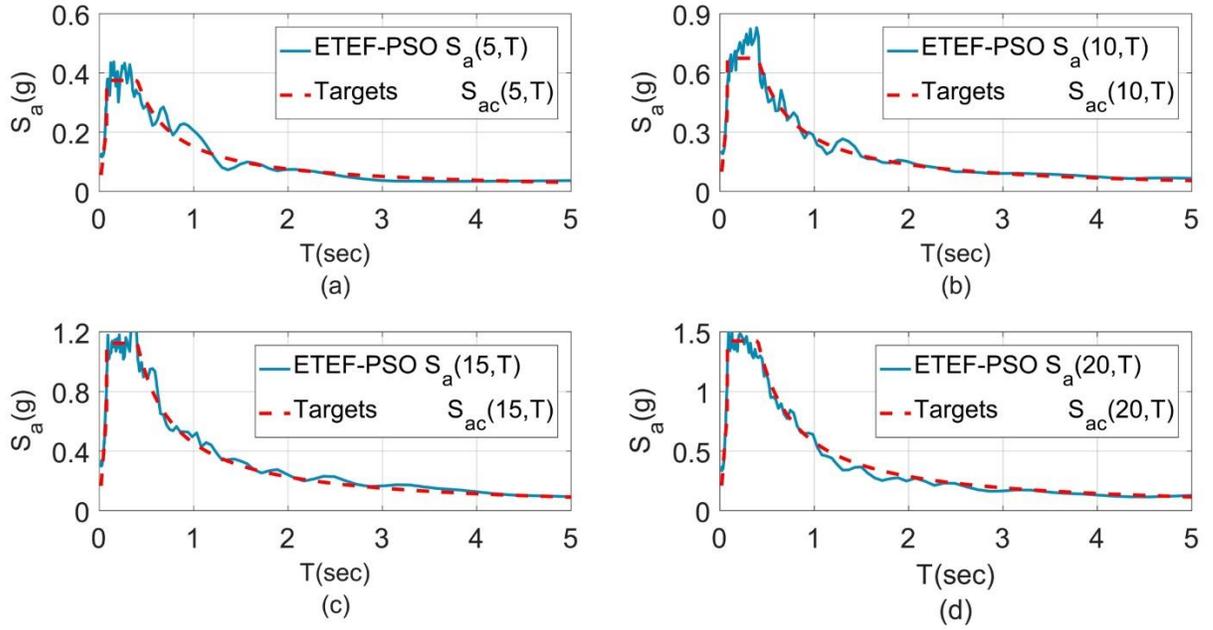

Figure 5. Response spectra of simulated ETEF-PSO against the target response spectra at (a) *t*=5sec, (b) *t*=10sec, (c) *t*=15sec, and (d) *t*=20sec

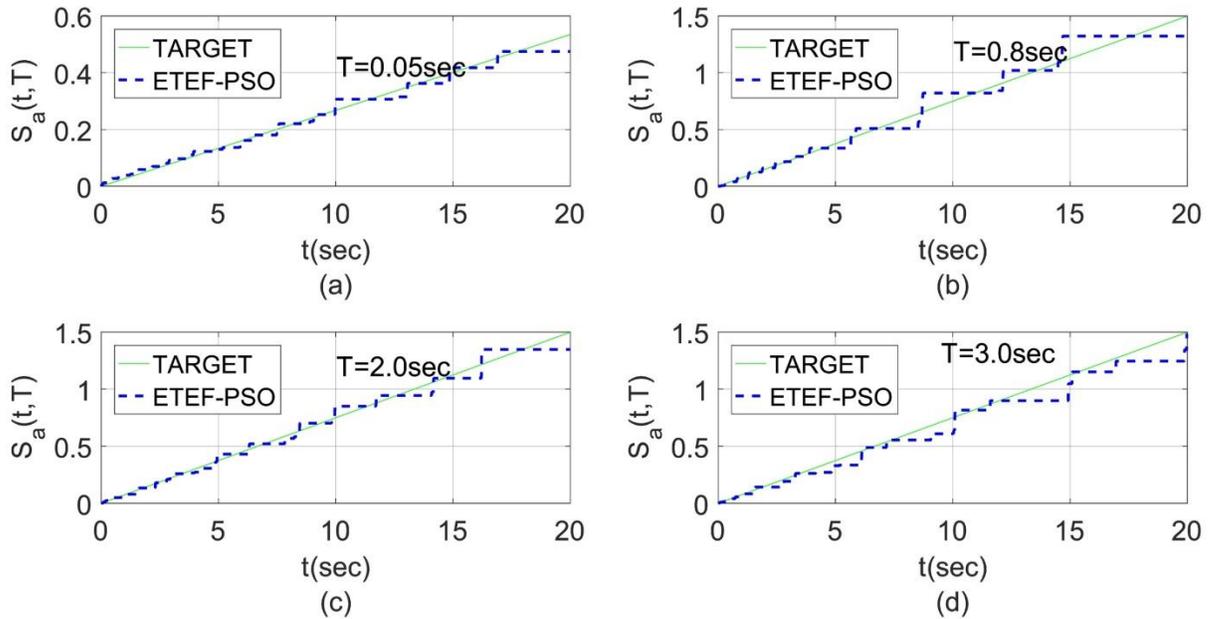

Figure 6. Targets vs. the time variation of response spectra in the simulated linear ETEF-PSO at different natural periods of (a) *T*=0.05sec, (b) *T*=0.8sec, (c) *T*=2.0sec, and (d) *T*=3.0sec

# 8 A Comparative evaluation between the proposed method and other existing procedures

In this section, results obtained using the proposed algorithm (the PSO-based ETEFs) are presented beside the outputs of the trust-region-reflective optimization framework as well as the ones related to the ICA method. In this case, several ETEFs are simulated based on the optimal parameters of the PSO-based algorithm for ETEF functions, which are provided in Table 2 along with their relevant values for the characteristics of best cost functions.

Table 3. Results of simulated ETEFs by using the PSO-based simulation with optimal parameters.

| Run number | Minimum objective function | | | Computational time (sec) | | | Number of function evaluations | | |
|---|---|---|---|---|---|---|---|---|---|
| | Values | Mean | Standard deviation | Values | Mean | Standard deviation | Values | Mean | Standard deviation |
| 1 | 795.80 | | | 307055 | | | 6000400 | | |
| 2 | 808.07 | | | 325894 | | | 6000400 | | |
| 3 | 795.45 | 804.66 | 23.62 | 307502 | 320764 | 32661 | 6000400 | 6000400 | 0 |
| 4 | 768.39 | | | 271996 | | | 6000400 | | |
| 5 | 831.31 | | | 350039 | | | 6000400 | | |
| 6 | 828.95 | | | 362098 | | | 6000400 | | |

Based on the information reflected in Table 3, a 59% improvement in the mean value, as well as a 98% decrease in the standard deviation, is gained for the minimum cost functions of the cases generated by the optimum parameters derived for the PSO algorithm for the ETEF simulations. In this case, the mean value of minimum cost function changes from 1966.13 to 804.62. Besides, the corresponding value for the standard deviation decreases to 23.62 while it is 1134.88 in cases evaluated in the parametric study (reported in Table 1). So, a lower standard deviation certainly leads to a lower number of required iterations which should be used to simulate a set of efficient PSO-based ETEFs

The trust-region algorithm is the current approach for simulating ETEFs, and it is considered as a classical optimization framework. For the comparison purposes, 6 ETEFs are simulated using the trust-region algorithm here in this study, which are based on 6 different initial guesses. The results are summarized in Table 4. Minimum objective function values, computational time and number of function evaluations are provided in this table. The mean and standard deviation of the minimum values for the objective function are 929 and 70, respectively. The mean and standard deviation of computational time are, respectively, 15443 sec and 372 sec. These values are used to compare the efficiency of trust-region algorithm and PSO in the following section.

Table 4. Results of simulated ETEFs by using trust-region reflective algorithm

| Random initial motion | Minimum objective function | Computational time (sec) | Number of function evaluations |
|---|---|---|---|
| | Values | Values | Values |
| $X_1$ | 971.83 | 15053 | 103626 |
| $X_2$ | 1013.62 | 16036 | 103626 |
| $X_3$ | 981.95 | 15738 | 103626 |
| $X_4$ | 832.914 | 15261 | 103626 |
| $X_5$ | 896.77 | 15188 | 103626 |
| $X_6$ | 879.51 | 15387 | 103626 |

The PSO is also compared with the imperialist competitive algorithm (ICA) which has been recently used for ETEFs simulating, where ICA is also categorized into a population-based evolutionary algorithm which is inspired by social-political behaviors. This algorithm also searches for the best possible solution from an initial population composed of components that may be qualified candidates as a start point. In order to compare the results from ICA with ones being simulated with the PSO, ICA is hired to simulate 6 ETEFs on the basis of 6 different initial motions generated by the optimal parameters reported in [17].

Table 5 compares the performance of trust-region reflective as a classical method with ICA and the PSO method in simulating ETEFs. It is interesting to find out that ETEFs simulated by ICA and PSO are about 11% and 14% more accurate, respectively, in terms of the mean value for the minimum cost function if they are compared to the ones produced through another employed method like the trust-region-reflective algorithm. However, ICA and the PSO require more than 50 times larger number of function evaluations in the trust-region reflective algorithm. But in terms of analysis time, the results indicated in Table 6 show that PSO is a rather faster algorithm if we compare it with ICA. The average analysis time required by the PSO is more than half of the value essential by ICA.

Table 5. Trust-region reflective vs. ICA and PSO in simulating ETEFs

| Value | Trust-region reflective | ICA | PSO |
|---|---|---|---|
| Best minimum objective function | 832.91 | 805.4 | 768.39 |
| Average minimum objective function | 929 | 832 | 804.66 |
| Worst minimum objective function | 1013.62 | 869.43 | 831.31 |
| Standard deviation of minimum objective function | 70 | 25 | 23.62 |
| Average analysis time (hour) | 4 | 202 | 89 |
| Average number of function evaluations | 103626 | 6596363 | 6000400 |

## 9 Discussion

In this section, we want to check the efficiency of newly generated ET excitations—the PSO-based ETEF—in structural dynamic analysis. In this case, the result obtained from a nonlinear time history analysis of an MDOF structure is taken as a benchmark for this evaluation. For comparison purposes, the case study MDOF model would be also subjected to an ETEF produced by an ICA-based algorithm [17]. Next, the results determined based on the aforementioned ETEFs are compared to the ones obtained through dynamic time history analyses performed under a set of selected ground motions. These ground motion records, as shown in Figure 7, are selected from a far-field dataset of FEMA P-695 [35]. To make a basis in predicting the engineering demand parameters (EDPs) of this study, an ETEF-PSO and an ETEF-ICA are generated using a unique target spectrum which is the median acceleration spectrum of the selected ground motions. These ETEFs are simulated based on the optimum parameters of the PSO-based simulation procedure and the corresponding factors reported for the ICA-based algorithm [17], respectively. Then, the response spectra of the selected ground motions along with the generated ETEFs—both the PSO-based and the ICA-based one—are scaled to 0.1g at the first mode period of vibration. It is worth to add that the roof displacement and inter-story drift ratios at different stories are considered as EDPs in this seismic response assessment.

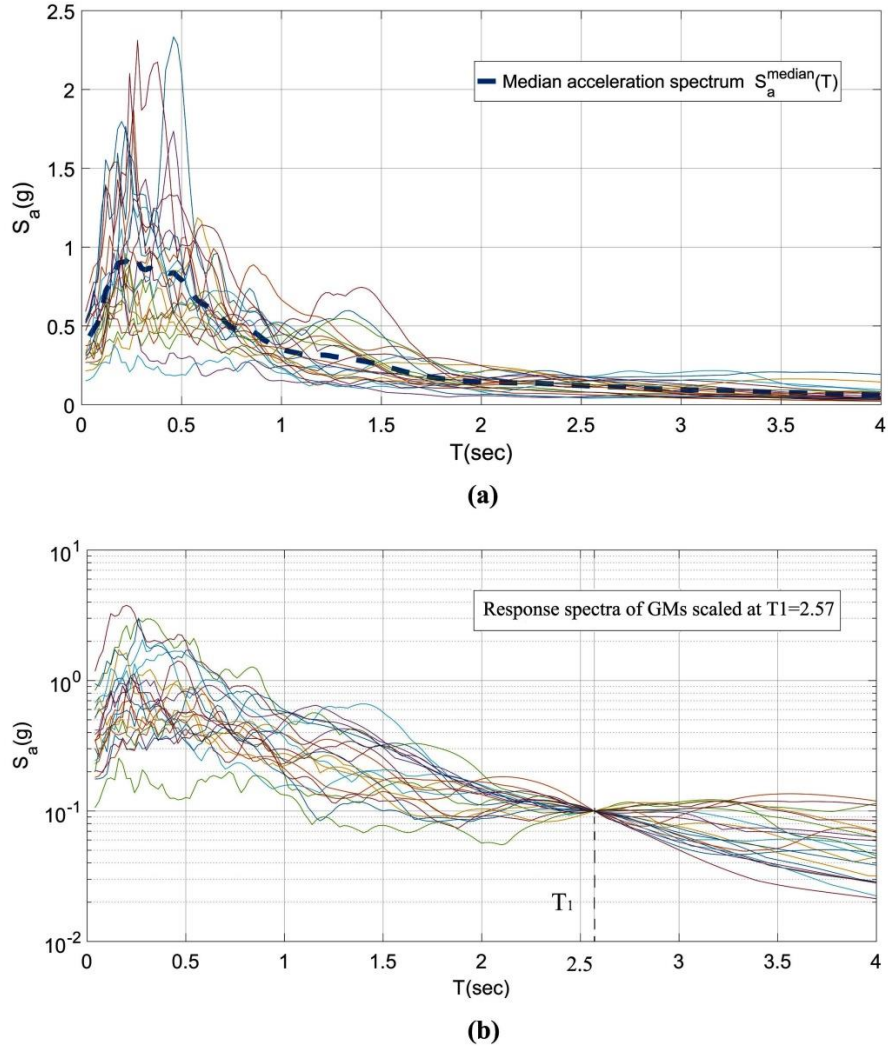

Figure 7. Acceleration spectra of the selected GMs: (a) response spectra of far-field set GMs of FEMA P-695, (b) response spectra of GMs scaled at T1

The well-known Bouc–Wen hysteretic model offered by Ma et al. [36] is utilized for structural modeling of this dynamic assessment, where the calibrating parameters of this employed hysteretic model are as follows: $p = 2$, $\psi = 0.05$, $\lambda = 0.5$, $q = 0.25$, $n = 1$, $\alpha = 0.04$, $\beta = 280$, $\gamma = 160$, $A = 1.0$, $\zeta_s = 0.99$, $\delta_\psi = 0.005$, $\delta_\nu = 0.002$, $\delta_\eta = 0.001$. Additionally, the Rayleigh damping is also considered to make this modeling become more realistic: $\mathbf{C} = a\mathbf{M} + b\mathbf{K}$, where a = 0.4602, b = 0.0041, $\mathbf{C}$, $\mathbf{K}$ and $\mathbf{M}$ are the damping matrix, stiffness matrix and mass matrix, respectively. Evidently, the material behavior used here for structural modeling is capable of including the degradations both for the strength and stiffness of the members. Configuration of the considered MDOF structure is presented in Figure 8. The modal characteristics of this structural model are also reported in Table 6.

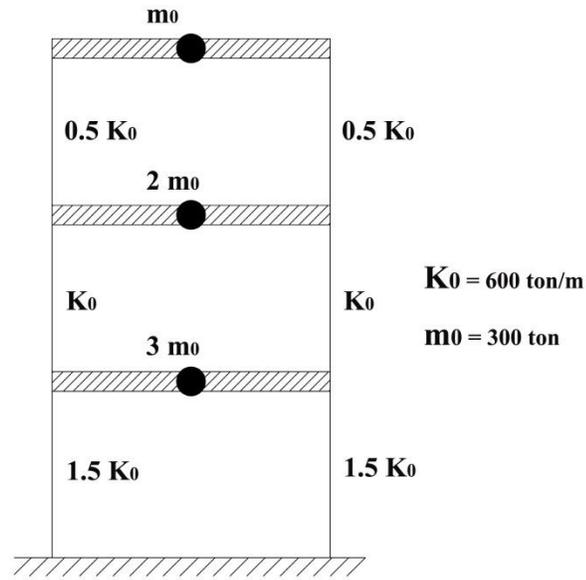

Figure 8. Configuration of the considered MDOF structure

Table 6. Characteristics of the considered MDOF model

| Mode number | Period (sec) | Participation factor (%) |
|---|---|---|
| 1 | 2.57 | 805.4 |
| 2 | 1.23 | 832 |
| 3 | 0.88 | 869.43 |

Results indicated in Table 7 demonstrate that the median EDPs of the employed ground motions can be successfully predicted by the applied PSO-based ETEF. In fact, the exactness of the results estimated by the PSO-based ETEF over the one related to the ICA-based ETEF is apparently recognizable. Since it is commonly accepted and known that higher modes substantially contribute to the nonlinear structural demands of MDOF structures, improved results can be attributed to the appropriate spectral matching at the higher modes of the ETEF produced by the PSO-based simulation method.

Table 7. EDP assessment using recorded GMs, ETEF-PSO and ETEF-ICA

| EDP | EDP assessment | | | Error percentage | |
|---|---|---|---|---|---|
| | ETEF-PSO | ETEF-ICA | GMs | ETEF-PSO (%) | ETEF-ICA (%) |
| First floor drift (%) | 0.045 | 0.041 | 0.046 | -2.2 | -10.9 |
| Second floor drift (%) | 0.046 | 0.042 | 0.048 | 2.1 | -12.5 |
| Third floor drift (%) | 0.056 | 0.049 | 0.057 | -1.8 | -14.0 |
| Roof displacement (cm) | 34.1 | 31.3 | 34.5 | -1.2 | -9.3 |

The matching quality of these PSO-based ETEFs is shown in Figure 9, where it can be readily acknowledged that matching is appropriately done at all considered periods. This matter highlights the significance of the devised algorithm standing for the ETEF spectral matching—or equivalently the objective function values achieved—that are utilized to fit the ETEFs to the preselected targets. Therefore, acceleration spectra compatibility over all required periods is necessary to achieve an acceptable seismic response assessment. This fact emphasizes the prominence of developing more efficient and accurate methodologies for generating new ETEFs.

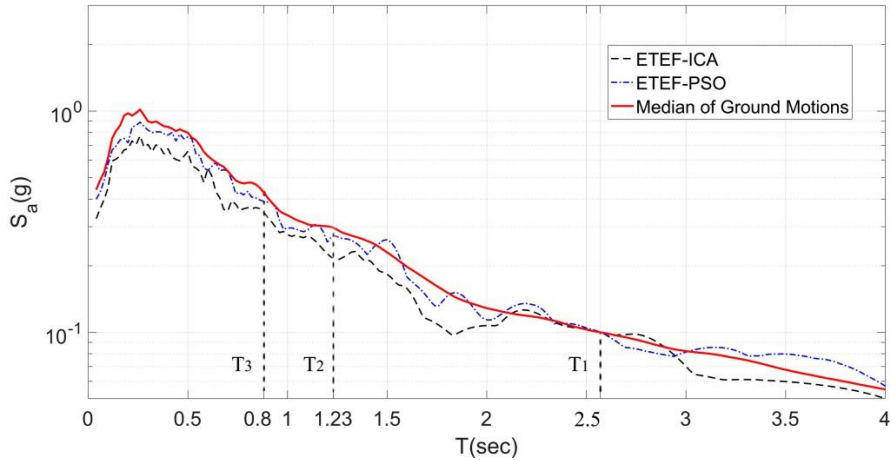

Figure 9. Comparison between the median spectrum of the selected GMs with the ones computed for ETEF-PSO and ETEF-ICA, all of which are scaled to 0.1g at the first mode period of the considered MDOF

## 10 Conclusions

Endurance time (ET) method is a dynamic analysis through which structural engineers are enabled to examine structures that are exposed to a set of predesigned increasing acceleration functions, the so-called Endurance Time Excitation Functions (ETEFs). When compared to a conventional procedure of dynamic time history analysis (DTHA), ET method demands a less computational time because ETEFs are designed in a way that they can deliver structural responses at several seismic levels in one single run. This is totally in contrast with the framework of a DTHA procedure through which structural responses would be available just in one seismic level after each analysis run. Since the fact that the intensifying shape of the ETEFs let us have structural demands at multiple seismic levels, it can be expected that accuracy of these responses would heavily depend on the exactness of such excitations—the ETEFs. As a result, there have been urgent demands for generating a set of more accurate ETEFs as a central part of the ET method in recent years. So ETEFs should be generated in a way that their dynamic characteristics would seem to get quite compatible with the recorded ground motions. In this case, a number of desired dynamic characteristics can be readily chosen as targets for ETEFs simulation using unconstrained nonlinear optimizations. However, two problems normally arise for such a simulation: the large number of decision variables and their highly correlated nature. In this paper, a PSO-based algorithm is proposed to cope with aforementioned difficulties. First, a discrete wavelet transform is employed to make a substantial reduction in the number of involved optimization variables. Next, a covariance matrix of variables generating an initial population is hired to properly control the highly correlational behavior of the existing variables. In the end, a set of ETEFs are generated using the PSO-based simulation of this paper. Statistical comparisons made between PSO-based ETEFs and the ones available show noticeable improvements. The main conclusions are collected below:

- Although it is shown that generation of the proposed PSO-based ETEFs is more time-consuming compared to other available methods, the efficiency of the ETEFs obtained by optimal parameters of the PSO-based algorithm is clearly demonstrated through the evidence which is based on their objective function and their exactness in nonlinear response assessments.

- Simulation of ETEFs with the reported optimal parameters of the proposed method delivers a group of ET excitations that are 59% more accurate than the ones generated with the arbitrary parameters of the PSO-based method. Accordingly, earthquake engineers can readily use the reported optimal parameters for simulation of new ETEFs for a location of interest as the desired site.

- A comparative evaluation between the dynamic characteristics of the PSO-based ETEFs and selected targets proves a perfect agreement, indicating the efficiency of the proposed method for the simulation of such ET excitations.

- The ETEFs produced with the proposed method are 4% and 14% more accurate if they are compared with outputs given by the trust-region-reflective and imperialist competitive algorithm, respectively.

- The produced ET excitations using the proposed method have about 65% and 6% less standard deviation when compared to the results produced by trust-region-reflective and imperialist competitive algorithm, respectively.

**Acknowledgment**

This study received financial support from the Iran's National Elites Foundation (INEF) under Grants No. AEIN-1100/967. The authors thank the anonymous reviewers whose comments have greatly improved this manuscript. The sincere efforts by the members of the HPC platform at the Sharif University of Technology are warmly appreciated because they provide very fast but reliable computing infrastructures for investigations that need demanding computational times. It is worth mentioning that the first two authors of this study have contributed equally to the work.

**Conflict of interest**

The authors declare that there is no conflict of interest regarding the publication of this article